\documentclass[11pt,a4paper]{article}

\usepackage[utf8]{inputenc}
\usepackage[T1]{fontenc}
\usepackage{lmodern}
\usepackage{amsmath,amssymb,bm}
\usepackage{siunitx}
\usepackage{booktabs}
\usepackage{graphicx}
\usepackage{geometry}
\usepackage{hyperref}
\usepackage{microtype}
\usepackage{authblk}
\usepackage{xcolor}

\usepackage{float}
\usepackage{placeins}

\hypersetup{
    colorlinks=true,
    linkcolor=blue!50!black,
    citecolor=blue!50!black,
    urlcolor=blue!50!black
}

\title{Efficient First-Principles Thermodynamic Modeling of Pressure–Composition–Temperature Diagrams in Multicomponent Alloys for Hydrogen Storage}

\author[1]{Anderson Silva Chaves}
\author[1]{Walter José Botta}
\author[1]{Guilherme Zepon}

\affil[1]{Department of Materials Engineering,
Federal University of São Carlos,
São Carlos, SP, Brazil}
\date{}

\begin{document}

\maketitle


\begin{abstract}
Predicting hydrogen sorption thermodynamics in multicomponent alloys remains challenging because chemical disorder and finite-temperature phase equilibria are computationally demanding. Here, we develop a first-principles-based thermodynamic model for pressure--composition--temperature (PCT) behavior in random multicomponent alloys. The framework combines density functional theory (DFT) formation enthalpies calculated using the coherent potential approximation Korringa--Kohn--Rostoker (CPA--KKR) method with a composition-dependent Gibbs free-energy formulation. By treating chemical disorder through an effective medium, CPA--KKR avoids large supercells required for random alloys, reducing the cost of generating energetic datasets. Enthalpies are combined with a NIST-based finite-temperature molecular hydrogen reference and configurational entropy for interstitial hydrogen occupation and site blocking. BCC, BCT, and FCC structures are considered for Ti--V--Nb--Cr-based alloys, including Al-containing compositions, and their relative stabilities are determined by DFT. Phase equilibria are obtained through numerical common-tangent constructions, yielding equilibrium plateau pressures and hydrogen-composition widths.
A single global parameter associated with the molecular hydrogen reference is determined using 22 experimental PCT conditions, which are also used to assess model predictions. The model can be evaluated at other thermodynamic conditions beyond this calibration and assessment dataset. Predicted plateau widths agree well with experiment: 95.5\% are within a factor of 1.5, with a mean absolute relative error of 26.1\%. For equilibrium pressures, 86.4\% of predictions lie within one order of magnitude of experimental values. This computationally efficient and physically transparent framework captures key PCT features and provides a baseline for broader alloy-space exploration.
\end{abstract}

\section{Introduction}
\label{sec1}

The increasing environmental impact associated with the
extensive use of fossil fuels, particularly greenhouse gas
emissions, has intensified the search for alternative energy
carriers capable of supporting a sustainable energy
infrastructure \cite{von2019application,pistidda2021solid}.
Among the proposed alternatives, hydrogen is particularly
attractive because of its high gravimetric energy content
and carbon-free utilization \cite{schlapbach2001hydrogen}.
However, the large-scale implementation of hydrogen
technologies remains strongly dependent on the development
of safe, efficient, and compact storage systems. Conventional
approaches based on compressed or cryogenic hydrogen involve
important limitations in terms of safety, volumetric energy
density, and energy consumption, motivating the development
of solid-state storage technologies \cite{Huot2010}.

Metal hydrides provide a promising route for solid-state
hydrogen storage because hydrogen can be reversibly
incorporated into interstitial sites of metallic lattices,
enabling high volumetric storage densities at moderate
pressures \cite{wu2025hydrogen}. The thermodynamic stability
of the hydrogenated phases determines the equilibrium
conditions for absorption and desorption and is therefore
a key factor in the design of practical storage materials.
Ideally, a hydrogen-storage alloy should combine reversible
hydrogen uptake near technologically relevant temperatures,
suitable equilibrium pressures, adequate storage capacity,
favorable kinetics, and long-term stability
\cite{dematteis2021substitutional}. These requirements are
difficult to satisfy simultaneously. Materials with high
hydrogen capacity may require elevated temperatures for
hydrogen release, whereas materials operating closer to
ambient conditions can exhibit insufficient capacity,
difficult activation, or unfavorable thermodynamic stability
\cite{dematteis2021substitutional,hirscher2020materials}.
Consequently, developing materials with tunable hydrogen
sorption thermodynamics remains an important challenge.

Multicomponent alloys derived from the high-entropy alloy
(HEA) concept~\cite{tsai2014high,miracle2017critical,george2019high} offer a particularly attractive materials-design
space for addressing this challenge. Such alloys contain
multiple principal elements that can form substitutional
solid solutions, with their chemical complexity providing
additional degrees of freedom for tailoring electronic
structure, lattice stability, and hydrogen--metal interactions
\cite{miracle2017critical}. This compositional flexibility
has motivated extensive investigation of multicomponent
alloys as hydrogen-storage materials and as a route toward
controlling hydride thermodynamics beyond the limitations
of conventional intermetallic compounds
\cite{marques2021review}. In particular, refractory BCC
alloys containing Ti, V, and Nb exhibit favorable
hydrogen-storage characteristics and can accommodate high
hydrogen concentrations \cite{zlotea2019hydrogen}. Their
thermodynamic behavior can be further tuned through additions
of elements such as Cr and Al, which modify the stability
of the hydrogenated states and can shift equilibrium
pressures toward conditions relevant to practical hydrogen
storage \cite{strozi2021effects,nygaard2019counting}.

The pressure--composition--temperature (PCT) diagram is
one of the most important descriptors of hydrogen absorption
in these materials. It provides direct information about
equilibrium hydrogen pressures, phase transformations,
and the composition intervals associated with two-phase
coexistence \cite{Huot2010,alvares2022modeling}. In
multicomponent alloys, however, the prediction of PCT
behavior is considerably more demanding than the calculation
of isolated formation energies. The equilibrium response
results from the competition among enthalpic, configurational entropy,
and hydrogen-reference contributions to the Gibbs free
energy, while the chemical disorder of the metallic
sublattice introduces an additional computational
challenge. Experimental PCT measurements remain
indispensable, but systematic experimental exploration
becomes increasingly difficult as the number of alloying
elements and possible compositions increases.
First-principles calculations provide an important route
for reducing this experimental burden. 

For the Ti--V--Nb-based alloy systems considered in this
work, the structural descriptions adopted to represent
hydrogen absorption are guided by phase-transformation
sequences reported in experimental studies of related
refractory multicomponent alloys. In particular, the
successive BCC, body-centered tetragonal (BCT), and
face-centered cubic (FCC) hydrogenated states provide
the structural basis for describing the two-step
hydrogenation behavior and the corresponding PCT plateaus
\cite{zlotea2019hydrogen,strozi2021effects,nygaard2019counting}.
Accordingly, the present thermodynamic framework considers
the sequence
\[
\mathrm{BCC}
\rightleftharpoons
\mathrm{BCT}
\rightleftharpoons
\mathrm{FCC},
\]
where the first equilibrium describes the BCC--BCT
transformation and the second describes the BCT--FCC
transformation. These structural descriptions are used
as phase-specific thermodynamic states in the model,
with their relative stability determined by the calculated
Gibbs free energies. Thus, the adopted sequence is
grounded in experimental observations reported in the
literature, while the equilibrium compositions and
pressures are obtained from the thermodynamic
calculations.

Density functional
theory (DFT)
\cite{hohenberg1964inhomogeneous,kohn1965self,kohn1999nobel}
has been widely used to investigate hydride formation
energies, electronic structures, and hydrogen site
stability
\cite{matar2010intermetallic,wu2025hydrogen,
dangwal2024machine,hannappel2025generalized}.
Conventional supercell-based DFT calculations, however,
become computationally expensive for random multicomponent alloys because sufficiently large cells are
required to represent chemical disorder and its associated
configurational space \cite{widom2018modeling,hannappel2025generalized}.
This limitation becomes particularly important when
calculations must be repeated over several alloy
compositions, hydrogen concentrations, and
crystallographic phases, as required for constructing
complete PCT diagrams.

Previous thermodynamic approaches to hydrogen storage in multicomponent alloys include the random site occupancy (RSO) model~\cite{Zepon2021}, the discrete site energy (DSE) model~\cite{PedrosoDSE}, and automated DFT--CALPHAD frameworks~\cite{Hannappel2025CALPHAD,Hannappel2026}. While these approaches account for different aspects of the thermodynamics of hydrogen absorption, they involve trade-offs between the description of local chemical environments and computational cost. In particular, explicit supercell calculations provide detailed information on local interstitial environments but become computationally demanding for increasingly complex alloys. The present work uses the coherent potential approximation (CPA)
\cite{Soven1967CPA,Velicky1968,gyorffy1972coherent}
combined with the Korringa--Kohn--Rostoker (KKR) method
\cite{Korringa1947,Kohn1954} to efficiently describe multicomponent alloy energetics and their hydrogen-storage thermodynamics.
The CPA-KKR method provides an efficient
first-principles framework for treating substitutional
chemical disorder without explicitly constructing
large supercells. Within the DFT framework, CPA--KKR
enables the electronic structure and energetic properties
of chemically disordered random alloys to be evaluated
at substantially reduced computational cost
\cite{ruban2008configurational}. This makes the approach
particularly suitable for generating the
composition-dependent energetic datasets required
for thermodynamic modeling of multicomponent hydrides.

Nevertheless, an efficient description of the disordered
electronic structure alone is not sufficient to predict
PCT behavior. A connection between the first-principles
energetics and finite-temperature phase equilibria must
also be established. In particular, the free-energy
landscape must account for hydrogen configurational
disorder and site blocking, as well as the
finite-temperature thermodynamics of the molecular
hydrogen reference state. The resulting phase equilibria
can then be obtained from the equality of chemical
potentials and the corresponding common-tangent
construction. An important challenge is to achieve
this without introducing a separate set of empirical
parameters for every alloy composition, since such
an approach would limit the transferability of the
resulting model and substantially reduce its
predictive value.
In this context, a thermodynamic formulation
provides a useful starting point for connecting
first-principles energetics to experimentally observed
PCT behavior. By retaining the essential energetic
and entropic contributions while excluding additional
finite-temperature terms, such as explicit hydrogen
and metallic-lattice vibrational free energies, the
individual role of the included contributions can be
examined within a clearly defined baseline model.
This approach also provides a reference against which
more elaborate thermodynamic descriptions can be
systematically evaluated.

In the present work, we develop a 
first-principles-based thermodynamic model for
predicting PCT behavior in random multicomponent
Ti--V--Nb-based alloys. The framework combines
CPA--KKR calculations with a phase-specific
Gibbs free-energy formulation for the BCC, BCT,
and FCC hydrogenated states. The microscopic energetic
input is provided by DFT-derived formation enthalpies
as a function of hydrogen concentration. These
enthalpies are combined with a finite-temperature
molecular hydrogen reference based on NIST
thermodynamic data and a configurational entropy
formulation describing hydrogen occupation of
interstitial sites, including a site-blocking treatment
for the BCC configuration. The resulting Gibbs
free-energy curves are used to determine the
BCC--BCT and BCT--FCC equilibria through numerical
common-tangent constructions. In this 
formulation, explicit hydrogen and metallic-lattice
vibrational free-energy contributions are not included.

The methodology is applied to seven
Ti--V--Nb--Cr-based multicomponent alloy compositions,
including systems with Al additions, over 22
experimental PCT conditions. A single global
parameter, $A_{\mathrm{298K}}$, is optimized to
align the molecular hydrogen reference with the
DFT energy scale. The calibration simultaneously
considers the experimental equilibrium pressures
and plateau widths of the second
BCT--FCC plateau. No alloy-specific or
temperature-specific fitting parameters are
introduced. The resulting model is assessed
globally by comparing its calculated second-plateau
pressures and widths with the corresponding
experimental values across all seven compositions
and 22 conditions.

The objective of this work is thus to determine whether
a compact thermodynamic description, supported by
computationally efficient first-principles
calculations, can generate the principal features
of experimental PCT diagrams across chemically
distinct multicomponent alloys using only one
global fitting parameter. Particular attention is
given to the distinction between the prediction
of equilibrium pressure and plateau width, since
these quantities provide complementary information
about the common-tangent construction and the
underlying Gibbs free-energy landscape. 

%
%
%
%
%
%
%
%
%


\section{Theoretical and Computational Framework}

\subsection{Para-equilibrium thermodynamic framework}
\label{sec:para_equilibrium}

Hydrogen absorption in the random multicomponent alloys (RMAs) considered in
this work is formulated within the para-equilibrium (PE) framework
\cite{alvares2022modeling}. This
description is appropriate for hydrogen-storage conditions in which the
metallic sublattice is effectively immobile on the timescale of hydrogen
redistribution, whereas hydrogen atoms remain mobile and can equilibrate
among the available phases. Under PE conditions, the overall metallic
composition is therefore preserved during hydrogen absorption, while the
hydrogen chemical potential is equilibrated between coexisting phases.

For a multicomponent alloy containing $n_{\mathrm M}$ metal atoms and
$n_{\mathrm H}$ hydrogen atoms, the hydrogen concentration is expressed as
the hydrogen-to-metal atomic ratio,

\begin{equation}
c_{\mathrm H}
=
\frac{n_{\mathrm H}}{n_{\mathrm M}}.
\label{eq:cH_definition}
\end{equation}

At a given temperature, the thermodynamic stability of each crystallographic
phase is described by its Gibbs free energy as a function of hydrogen
concentration,

\begin{equation}
G^{\alpha}(c_{\mathrm H},T)
=
H^{\alpha}(c_{\mathrm H},T)
-
T S^{\alpha}(c_{\mathrm H},T),
\label{eq:G_general_PE}
\end{equation}

where $\alpha$ denotes the crystallographic phase. In the present baseline
formulation, the free energy is constructed from the DFT-derived energetic
description, the thermodynamics of the molecular hydrogen reference, and the
configurational entropy associated with hydrogen occupation of interstitial
sites.
The hydrogen chemical potential in a solid phase is obtained from the
composition derivative of the Gibbs free energy,

\begin{equation}
\mu_{\mathrm H}^{\alpha}
=
\left(
\frac{\partial G^{\alpha}}
{\partial c_{\mathrm H}}
\right)_T .
\label{eq:muH}
\end{equation}

The corresponding chemical potential of hydrogen in the molecular hydrogen
reservoir is

\begin{equation}
\mu_{\mathrm H}^{g}
=
\frac{1}{2}RT
\ln
\left(
\frac{P_{\mathrm H_2}}{P^0}
\right),
\label{eq:muH_gas}
\end{equation}

where $P^0$ is the standard pressure. 
Equilibrium between the hydrogen reservoir and the hydrogen-containing solid requires equality of the hydrogen chemical potentials:

\begin{equation}
\mu_{\mathrm H}^{\alpha}
=
\mu_{\mathrm H}^{g}.
\label{eq:muH_equilibrium}
\end{equation}

Under para-equilibrium, the relative proportions of the metallic elements are constrained, and their individual chemical potentials are not considered independently. Instead, the metallic sublattice is treated as a single effective component, whose chemical potential is given by

\begin{equation}
\mu_M = G-c_{\mathrm H}\mu_{\mathrm H},
\label{eq:muM}
\end{equation}

where \(G\) is the Gibbs free energy per mole of metal atoms and \(c_{\mathrm H}\) is the hydrogen-to-metal ratio. 








\subsection{Random multicomponent alloy structural description}
\label{sec:structural_models}

The metallic sublattice of each alloy is modeled as a random multicomponent
alloy (RMA), in which the constituent elements occupy the sites of a common
crystallographic lattice according to their prescribed atomic fractions. The
metallic sublattice is therefore treated as a substitutionally disordered
solid solution rather than as an ordered intermetallic compound.
Three crystallographic descriptions are considered in the thermodynamic
framework: body-centered cubic (BCC), body-centered tetragonal (BCT), and
face-centered cubic (FCC). These configurations provide the structural
branches required to describe the hydrogen absorption process. Their relative
thermodynamic stability is not imposed a priori, but follows from the
corresponding free-energy curves. 
Hydrogen
concentration is varied continuously within each phase, while the chemical
composition of the metallic sublattice remains unchanged.

\subsection{\label{section:structures}Hydride structures and hydrogen-site occupation}

Hydrogen absorption was represented using distinct crystallographic
configurations of the metallic sublattice, namely BCC, BCT, and FCC. 
The structural configurations provide the free-energy
branches used subsequently to determine the equilibrium phase relations.
For the BCC configuration, hydrogen occupies a randomly populated
tetrahedral sites. In the configurational model, the number of
interstitial sites available per metal atom is represented by the 
parameter $
\theta_{\mathrm{BCC}}=6.
$
The corresponding hydrogen-site occupancy is therefore defined as

\begin{equation}\label{eq:fi}
f_i=\frac{c_{\mathrm H}}{\theta}.
\end{equation}

Hydrogen--hydrogen interactions in this randomly occupied configuration are
represented through a concentration-dependent site-blocking fraction. This
effect is described by the JMAK-type formulation introduced in
Section~\ref{configurational_entropy}. 
For the BCT configurations, hydrogen ordering is represented through a
tetragonal structural description derived from the BCC lattice. Both
tetrahedral and octahedral hydrogen configurations are considered in the
structural calculations, corresponding to the $I\bar{4}m2$ (No.~119) and
$I4/mmm$ (No.~139) space groups, respectively. In the configurational
thermodynamic model, these ordered BCT configurations are represented by the site parameter
$
\theta_{\mathrm{BCT}}=1.
$
The same effective configurational treatment is used for the two BCT
structural variants, while their energetic properties are
determined independently from the corresponding first-principles
calculations.
At high hydrogen concentrations, an FCC hydride configuration with space
group $Fm\bar{3}m$ (No.~225) is considered. Hydrogen occupies tetrahedral interstitial
sites, and the effective number of sites used in the configurational model is $
\theta_{\mathrm{FCC}}=2.
$

\subsection{Alloy systems considered}

The model was applied to seven Ti--V--Nb-based
random multicomponent alloys. In the notation adopted
throughout this work, $\mathrm{TiVNb}_{x}\mathrm{Cr}_{y}$
represents an alloy in which Ti, V, and Nb contribute
equally to the fraction $x$, while Cr accounts for
the fraction $y$. Similarly,
$\mathrm{TiVNb}_{x}\mathrm{CrAl}_{y}$ denotes alloys
in which Ti, V, and Nb contribute equally to the
fraction $x$, while Cr and Al share the additional
fraction $y$ equally. The investigated compositions
are $\mathrm{TiVNb}_{80}\mathrm{CrAl}_{20}$\cite{Ferreira2023TiVNbCrAl}, 
$\mathrm{TiVNb}_{90}\mathrm{CrAl}_{10}$\cite{Ferreira2023TiVNbCrAl},
$\mathrm{TiVNb}_{85}\mathrm{Cr}_{15}$\cite{Silva2021TiVNbVEC}, 
$\mathrm{TiVNb}_{75}\mathrm{Cr}_{25}$\cite{Nygard2021TiVCrNb},
$\mathrm{TiVNb}_{70}\mathrm{Cr}_{30}$\cite{Strozi2023TiVNbCr},
$\mathrm{TiVNb}_{67}\mathrm{Cr}_{33}$\cite{Silva2023RoomTemperatureReversibility},
$\mathrm{TiVNb}_{65}\mathrm{Cr}_{35}$\cite{Strozi2023TiVNbCr}.

These compositions provide a systematic variation
of the Cr and Cr--Al contents within the Ti--V--Nb
matrix. For each alloy, the BCC, BCT, and FCC
structural descriptions are considered over the
hydrogen concentration range required for the
thermodynamic calculations. The corresponding
phase-dependent energetic and configurational
contributions are used to construct the Gibbs
free-energy curves and determine the BCC--BCT
and BCT--FCC equilibria through the common-tangent
construction.

\section{Thermodynamic formulation}

The equilibrium pressure--composition--temperature
(PCT) behavior was described using a 
thermodynamic model, in which the Gibbs free
energy is constructed separately for the BCC,
body-centered tetragonal (BCT), and face-centered
cubic (FCC) phases. The formulation combines
DFT-derived formation enthalpies with the
finite-temperature thermodynamic properties of
molecular hydrogen and the configurational entropy
associated with hydrogen occupation of interstitial
sites.

For a given phase $\alpha$, hydrogen concentration
$c_{\mathrm H}$, and temperature $T$, the Gibbs
free energy is expressed as

\begin{equation}
\begin{aligned}
G_{^{\alpha}}(c_{\mathrm H},T)
={}&
H_{\mathrm{DFT}}^{\alpha}(c_{\mathrm H})
\\
&-
\frac{c_{\mathrm H}}{2}
\left[
A_{\mathrm{298K}}
+
\Delta H_{\mathrm H_2}^{\mathrm{NIST}}(T)
-
T S_{\mathrm H_2}^{\mathrm{NIST}}(T)
\right]
\\
&-
T S_{\mathrm{conf}}^{\alpha}(c_{\mathrm H}).
\end{aligned}
\label{eq:M0_free_energy}
\end{equation}

Here, $H_{\mathrm{DFT}}^{\alpha}$ is the
phase-dependent formation enthalpy obtained from
first-principles calculations, referenced to the BCC structure, which is taken as reference, and the
molecular hydrogen reference. The term
$\Delta H_{\mathrm H_2}^{\mathrm{NIST}}$ represents
the temperature-dependent increment in the
molecular hydrogen enthalpy relative to
$T_{\mathrm{ref}}=298.15$~K, while
$S_{\mathrm H_2}^{\mathrm{NIST}}$ is the
corresponding molar entropy obtained from NIST.
The parameter $A_{\mathrm{298K}}$ accounts for
the residual energetic correction between the
DFT-based formation enthalpies and the
finite-temperature hydrogen reference.
Finally, $S_{\mathrm{conf}}^{\alpha}$ describes
the configurational entropy associated with
hydrogen occupation of the available interstitial
sites, including the site-blocking treatment
for BCC.
All energetic contributions are expressed in
$\mathrm{kJ\,mol^{-1}}$ of metal atoms, and
entropy contributions in
$\mathrm{kJ\,mol^{-1}\,K^{-1}}$ of metal atoms.
The factor $c_{\mathrm H}/2$ accounts for the
molecular nature of the hydrogen reservoir.

The model contains only one adjustable parameter,
$A_{\mathrm{298K}}$, optimized globally using the
experimental PCT data from the 22 investigated
conditions. The equilibrium pressures and plateau
widths are not fitted independently, but emerge
from the phase-dependent Gibbs free-energy curves
through the common-tangent construction.

\subsection{DFT enthalpy and hydrogen reference correction}

The first-principles calculations provide the
phase- and composition-dependent energetic input
to the thermodynamic model. Electronic structure
calculations were performed using the all-electron
Korringa--Kohn--Rostoker (KKR)\cite{Korringa1947KKR,Kohn1954KKR} Green's-function
method within density functional theory (DFT),
as implemented in the \textit{AkaiKKR}
(Machikaneyama) code\cite{Akai1989KKRCPA}. Spin-polarized calculations
were carried out for the hydrogen-free
multicomponent alloys and their corresponding
hydrogenated structures, considering the BCC,
BCT, and FCC phases over the hydrogen concentration
range required for the PCT calculations.

The KKR Green's-function formalism is based on
multiple scattering theory, in which the electronic
structure is obtained directly from the
one-electron Green's function associated with
the Kohn--Sham equation. 
The electron density, $\rho(\mathbf r)$, is evaluated from the
retarded Green's function by integrating its imaginary part up
to the Fermi energy, $E_{\mathrm F}$. Within the KKR formalism,
the Green's function is decomposed into a single-site contribution
and a multiple-scattering term, yielding

\begin{equation}
\begin{aligned}
\rho(\mathbf r)
={}&
\sum_i |\phi_i(\mathbf r)|^2
-\frac{1}{\pi}
\operatorname{Im}
\lim_{\mathbf r'\rightarrow\mathbf r}
\int_{-\infty}^{E_{\mathrm F}}
\left[
G^{s}(\mathbf r,\mathbf r',\epsilon)
+
G^{m}(\mathbf r,\mathbf r',\epsilon)
\right]d\epsilon ,
\end{aligned}
\label{eq:green_density}
\end{equation}

where $\phi_i(\mathbf r)$ denotes the core-state wave functions,
$G^{s}(\mathbf r,\mathbf r',\epsilon)$ represents the
single-site contribution associated with the scattering
potential, and $G^{m}(\mathbf r,\mathbf r',\epsilon)$ accounts
for multiple-scattering processes between atomic sites.
The limit $\mathbf r'\rightarrow\mathbf r$ yields the
site-diagonal Green's function required for the evaluation
of the electron density.
The integration
is performed along a contour in the complex
energy plane, providing a numerically stable
evaluation of the electronic density and
ground-state properties.
The energy contour extended over a width of
$2.0$~Ry below the Fermi energy, ensuring the
inclusion of all valence bands. A total of 71
energy points were employed along a semicircular
contour in the upper half of the complex energy
plane. A small imaginary component of
$\eta=10^{-4}$~Ry was added to the Fermi energy
to avoid numerical singularities associated with
the branch cut along the real energy axis.
An angular momentum cutoff of $l_{\mathrm{max}}=3$
was adopted for the expansion of the Green's
function. Brillouin zone integrations were
performed using 256 $k$-sampling points in the
first irreducible Brillouin zone.
Exchange--correlation effects were described
using the Perdew--Burke--Ernzerhof (PBE)
generalized gradient approximation\cite{Perdew1996PBE}. Relativistic
effects were included within the scalar-relativistic
approximation\cite{Akai1989KKRCPA}, and the crystal potential was
treated using the atomic-sphere approximation
(ASA)\cite{Temmerman1978KKRASA}. These approximations were consistently
applied to all investigated metallic and
hydrogenated structures.

The multicomponent alloys were modeled assuming
completely random substitutional atomic
arrangements. Chemical disorder was treated
within the coherent potential approximation
(CPA)\cite{Soven1967CPA,Velicky1968,gyorffy1972coherent}, in which the scattering effects arising
from the different atomic species are represented
by an effective medium characterized by a
self-consistent, energy-dependent coherent
$t$-matrix. This approach enables the
configurationally averaged electronic properties
of disordered alloys to be evaluated without
explicitly constructing large configuration-
dependent supercells, substantially reducing
the computational time  associated with the
systematic investigation of different alloy
compositions and hydrogen concentrations.
The self-consistent field  calculations
were initialized using a superposition of atomic
potentials and atomic $t$-matrices, approximated
using muffin-tin potentials within the ASA.
During the initial iterations, the Chebyshev
iterative mixing scheme was employed to handle
large root-mean-square (RMS) errors and charge
oscillations\cite{Akai1985Chebyshev}. Once numerical stability was
achieved, the mixing procedure was switched
to Broyden's second method to accelerate
convergence\cite{Johnson1988Broyden}. The potential mixing parameter
was set to 0.0036, and convergence was
achieved when the RMS error in the total
potential decreased below $10^{-6}$~Ry.
The same numerical parameters and convergence
criteria were adopted for all investigated
systems.

For each alloy composition, crystallographic
phase, and hydrogen concentration, the total
energy was evaluated as a function of volume.
The equilibrium structural energy was determined
from the minimum of the corresponding
energy--volume relation. 
The equation of state, based on the third-order Birch--Murnaghan formalism 
\cite{Birch1947,Murnaghan1937}, and the corresponding fitting parameters are described in the Supplementary Information. The resulting
equilibrium total energies constitute the
first-principles input for the formation
enthalpy calculations.

\subsubsection{Computational Efficiency of the KKR-CPA Approach}

The computational efficiency of the KKR-CPA method is particularly advantageous for investigating chemically disordered multicomponent alloys and their hydrides. Unlike conventional supercell-based DFT calculations, which require explicit atomic configurations to represent chemical disorder, the KKR-CPA method describes the configurationally averaged electronic structure through an effective medium. Consequently, for BCC and FCC structures, the chemical disorder can be treated using a primitive unit cell containing only one crystallographic metal site, irrespective of the number of alloying elements or their concentrations. In contrast, supercell approaches require a sufficiently large number of metal atoms to reproduce the desired composition and configurational correlations, often involving special quasirandom structures (SQS)\cite{Zunger1990SQS,Wei1990SQS}. The computational cost of conventional DFT calculations consequently increases substantially with the number of atoms in the supercell\cite{Hasnip2014DFTsolidstate}, making systematic investigations of multiple compositions and hydrogen concentrations computationally demanding. The efficiency of AkaiKKR for this type of problem has been demonstrated by Fukushima et al., who performed high-throughput calculations for 147,630 quaternary alloy systems using the KKR-CPA approach \cite{Fukushima2022AkaiKKR}. 

This advantage becomes even more relevant when hydrogen-containing structures are considered. In supercell calculations, hydrogen atoms must be explicitly introduced into the interstitial sites, increasing the number of atoms in the simulation cell and introducing additional configurational degrees of freedom associated with hydrogen occupation and local metal environments. Within the KKR-CPA framework, substitutional disorder on the metallic sublattice can instead be treated statistically, while hydrogen occupation can be incorporated through an appropriate effective-medium description of the interstitial sublattice. This substantially reduces the computational effort required to investigate multiple alloy compositions and hydrogen concentrations. Nevertheless, the single-site CPA approximation does not explicitly describe local lattice distortions or short-range chemical correlations, which can be investigated more accurately using explicit supercell calculations. Thus, KKR-CPA provides a computationally efficient approach for systematic thermodynamic screening of multicomponent alloys and their hydrides, whereas supercell-based DFT calculations remain valuable for examining specific local atomic configurations and relaxation effects \cite{song2017local,Karabin2022CPAComparison,Rittiruam2022AkaiKKR}.

\subsubsection{Hydrogen formation enthalpy and energy reference}

The hydrogen formation enthalpies were calculated using the total energy of the hydrogen-free alloy in its most stable BCC configuration as the common metallic reference. This choice sets the enthalpy of the BCC phase to zero at zero hydrogen concentration, while preserving the energetic difference between the hydrogen-free BCC and FCC structures.
An important structural aspect is that, in the zero-hydrogen limit ($c_{\mathrm H}=0$), both BCT configurations considered in this work, namely the tetrahedral and octahedral variants, correspond exactly to the BCC structure. Therefore, they have the same total energy as the hydrogen-free BCC configuration and, consequently, their formation enthalpies are also zero at $c_{\mathrm H}=0$. The distinction between these BCT configurations and the BCC structure emerges only upon hydrogen insertion.

For a hydrogenated configuration belonging to phase $\alpha$, the formation enthalpy is defined as
\begin{equation}
H_{\mathrm{DFT}}^\alpha(c_{\mathrm H}) =
E_{\mathrm{tot}}(M\mathrm H_{c_{\mathrm H}}^\alpha)
-
E_{\mathrm{tot}}(M^{\mathrm{BCC}})
-
\frac{c_{\mathrm H}}{2}E_{\mathrm{ref}}(\mathrm H_2),
\label{eq:formation_enthalpy}
\end{equation}
where $E_{\mathrm{tot}}(M\mathrm H_{c_{\mathrm H}}^\alpha)$ is the total energy of the hydrogenated alloy in phase $\alpha$, $E_{\mathrm{tot}}(M^{\mathrm{BCC}})$ is the total energy of the hydrogen-free alloy in its most stable BCC configuration, and $E_{\mathrm{ref}}(\mathrm H_2)$ is the reference energy per hydrogen molecule. The factor $c_{\mathrm H}/2$ accounts for the number of hydrogen molecules per metal atom.
Thus, the FCC formation-enthalpy curve includes the energy difference between the hydrogen-free FCC and BCC structures as a constant vertical offset. This offset is retained at every hydrogen concentration, ensuring that the BCC and FCC enthalpy curves are expressed on the same energetic reference.

Since a direct calculation of the isolated
$\mathrm H_2$ molecule is not available within
the present KKR--CPA implementation (as the best of our knowledge), the
effective molecular hydrogen reference was
established using the experimental formation
enthalpy of FCC $\mathrm{NbH}_2$. The experimental
value adopted was \cite{Fukai2005MetalHydrogen}

\begin{equation}
\Delta H_{\mathrm f}^{\mathrm{exp}}(\mathrm{NbH}_2)
=
-40.5238\;
\mathrm{kJ\,mol^{-1}\,H_2}.
\label{eq:NbH2_exp}
\end{equation}

The effective molecular hydrogen reference was
then determined from the calculated total
energies of hydrogen-free Nb and FCC
$\mathrm{NbH}_2$:

\begin{equation}
E_{\mathrm{ref}}(\mathrm H_2)
=
2\left[
E_{\mathrm{tot}}(\mathrm{NbH}_2)
-
E_{\mathrm{tot}}(\mathrm{Nb})
-
\Delta H_{\mathrm f}^{\mathrm{exp}}(\mathrm{NbH}_2)
\right].
\label{eq:H2_reference}
\end{equation}

The resulting $E_{\mathrm{ref}}(\mathrm H_2)$
was used consistently in the formation enthalpy
calculations of all investigated alloy
compositions, phases, and hydrogen concentrations.
Thus, a common molecular hydrogen reference is
established before the construction of the
thermodynamic free-energy curves.

\subsubsection{Temperature-dependent hydrogen reference correction}

The DFT formation enthalpies are obtained from
electronic ground-state energies at $0$~K,
whereas the thermodynamic properties of molecular
hydrogen vary with temperature. To incorporate
this temperature dependence, the molecular
hydrogen enthalpy and entropy were obtained from
the NIST Shomate representation \cite{Chase1998JANAF,NISTChemistryWebBook}.

Defining

\begin{equation}
t=\frac{T}{1000},
\end{equation}

the molar enthalpy of molecular hydrogen is
expressed as

\begin{equation}
\begin{aligned}
H_{\mathrm H_2}^{\mathrm{NIST}}(T)
={}&
A_{\mathrm S}t
+\frac{B_{\mathrm S}t^2}{2}
+\frac{C_{\mathrm S}t^3}{3}
+\frac{D_{\mathrm S}t^4}{4}\\
&-\frac{E_{\mathrm S}}{t}
+F_{\mathrm S}-H_{\mathrm S},
\end{aligned}
\label{eq:shomate_H2}
\end{equation}

while the corresponding molar entropy is

\begin{equation}
\begin{aligned}
S_{\mathrm H_2}^{\mathrm{NIST}}(T)
={}&
A_{\mathrm S}\ln t
+B_{\mathrm S}t
+\frac{C_{\mathrm S}t^2}{2}
+\frac{D_{\mathrm S}t^3}{3}\\
&-\frac{E_{\mathrm S}}{2t^2}
+G_{\mathrm S}.
\end{aligned}
\label{eq:shomate_S_H2}
\end{equation}
where 
$A_{\mathrm S} = 33.066178$,
$B_{\mathrm S} = -11.363417$,
$C_{\mathrm S} = 11.432816$,
$D_{\mathrm S} = -2.772874$,
$E_{\mathrm S} = -0.158558$,
$F_{\mathrm S} = -9.980797$,
$G_{\mathrm S} = 172.707974$ and 
$H_{\mathrm S} = 0$ are the NIST Shomate coefficients.
The resulting molar enthalpy is expressed in
$\mathrm{kJ\,mol^{-1}}$, whereas the molar entropy
is initially obtained in $\mathrm{J\,mol^{-1}\,K^{-1}}$
and subsequently converted to
$\mathrm{kJ\,mol^{-1}\,K^{-1}}$ to ensure consistency
with the energetic terms used in the free-energy model. 
The temperature-dependent enthalpy increment
relative to $T_{\mathrm{ref}}=298.15$~K is

\begin{equation}
\Delta H_{\mathrm H_2}^{\mathrm{NIST}}(T)
=
H_{\mathrm H_2}^{\mathrm{NIST}}(T)
-
H_{\mathrm H_2}^{\mathrm{NIST}}(T_{\mathrm{ref}}).
\label{eq:deltaH_H2}
\end{equation}

The molecular hydrogen contribution to the
Gibbs free energy is therefore

\begin{equation}
G_{\mathrm H_2,\mathrm{term}}(c_{\mathrm H},T)
=
-\frac{c_{\mathrm H}}{2}
\left[
A_{\mathrm{298K}}
+
\Delta H_{\mathrm H_2}^{\mathrm{NIST}}(T)
-
T S_{\mathrm H_2}^{\mathrm{NIST}}(T)
\right].
\label{eq:G_H2_term1}
\end{equation}

The single parameter $A_{\mathrm{298K}}$ is
determined by minimizing the discrepancy
between the calculated and experimental
equilibrium pressures over the 22 investigated
PCT conditions. Once optimized, the same
value is applied to every alloy composition
and temperature. No additional scaling or
alloy-dependent fitting parameters are
introduced.

\subsection{\label{configurational_entropy}Configurational entropy and site blocking}

The configurational contribution accounts for the
statistical distribution of hydrogen atoms and vacancies
over the available interstitial sites of each
crystallographic structure. Since the number of
interstitial sites per metal atom depends on the host
structure, the configurational entropy is formulated
using a phase-dependent number of available sites,
together with a site-blocking description for the
randomly occupied BCC interstitial sublattice.

For a given structural model containing $\theta$
interstitial sites per metal atom, the fractional
hydrogen occupancy is defined by Equation~\ref{eq:fi}. 
As described in Section~\ref{section:structures},
the values of $\theta$ adopted for the structural
models are

\begin{equation}
\theta =
\begin{cases}
6, & \mathrm{random\ interstitial\ BCC},\\
1, & \mathrm{ordered\ BCT\ (tetrahedral)},\\
1, & \mathrm{ordered\ BCT\ (octahedral)},\\
2, & \mathrm{FCC\ hydride}.
\end{cases}
\label{eq:theta_values}
\end{equation}

To account phenomenologically for hydrogen--hydrogen
interactions, a site-blocking model is introduced
through the fraction of interstitial sites rendered
unavailable for further hydrogen occupation, denoted
by $f_b$. The fraction of accessible sites is therefore

\begin{equation}
A=1-f_b.
\label{eq:available_sites}
\end{equation}

The effective hydrogen occupancy of the remaining
accessible sites is defined as

\begin{equation}
x=\frac{f_i}{1-f_b}
=\frac{f_i}{A}.
\label{eq:effective_occupancy}
\end{equation}

The corresponding configurational entropy is expressed
as

\begin{equation}
S_{\mathrm{conf}}^{\alpha}(c_{\mathrm H})
=
-R\theta A
\left[
x\ln x
+
(1-x)\ln(1-x)
\right],
\label{eq:Sconf_general}
\end{equation}

where $R$ is the universal gas constant. The factor
$\theta A$ represents the number of accessible
interstitial sites per metal atom. This expression
reduces to the conventional ideal configurational
entropy when no sites are blocked ($f_b=0$), for which
$A=1$ and $x=f_i$.

\subsubsection{JMAK site-blocking description}

For the randomly occupied BCC interstitial sublattice,
hydrogen--hydrogen interactions are represented
phenomenologically through a concentration-dependent
site-blocking fraction based on a
Johnson--Mehl--Avrami--Kolmogorov (JMAK)-type expression, as proposed by Pedroso et al. \cite{Pedroso2024JMAKSiteBlocking}:

\begin{equation}
f_b=1-\exp(-N_b f_i),
\label{eq:JMAK_blocking}
\end{equation}

where $N_b$ is an effective site-blocking parameter
controlling the rate at which accessible interstitial
sites become unavailable as hydrogen occupancy
increases. In the present implementation, following the approach proposed by Pedroso et al. \cite{Pedroso2024JMAKSiteBlocking}, $N_{\mathrm{b}}=14$ is adopted for the randomly occupied interstitial BCC configuration. The corresponding
accessible-site fraction is therefore

\begin{equation}
A=1-f_b=\exp(-N_b f_i).
\label{eq:JMAK_available}
\end{equation}

The JMAK treatment provides a continuous reduction
of the accessible interstitial sublattice with
increasing hydrogen concentration. In contrast,
no additional site blocking is introduced for the
ordered BCT and FCC configurations. This is
implemented by setting $N_b=0$, resulting in
$f_b=0$ and $A=1$. Equation~\eqref{eq:Sconf_general}
then reduces to

\begin{equation}
S_{\mathrm{conf}}^{\alpha}(c_{\mathrm H})
=
-R\theta
\left[
f_i\ln f_i
+
(1-f_i)\ln(1-f_i)
\right].
\label{eq:Sconf_no_blocking}
\end{equation}

Thus, a common configurational entropy formalism
is employed for all structural models, while the
JMAK site-blocking contribution is activated
exclusively for the randomly occupied BCC
interstitial configuration. The BCT octahedral
structural description, when considered in the
energetic analysis, retains its corresponding
site multiplicity but does not constitute an
additional competing phase in the PCT equilibrium
sequence.


\section{Common-tangent construction and equilibrium pressure}

Once the phase-specific Gibbs free-energy functions
$G^{\alpha}(c_{\mathrm H},T)$ have been constructed, the
pressure--composition--temperature (PCT) response follows from the resulting
free-energy landscape. At a fixed temperature, the hydrogen chemical
potential is obtained from the derivative of the Gibbs free energy with
respect to the hydrogen concentration given by Equation~(\ref{eq:muH})
and the corresponding
chemical potential of the metallic sublattice is given by Equation~(\ref{eq:muM}). 
Consequently, the equilibrium conditions can be
determined directly from the phase-specific Gibbs free-energy curves.

\subsection{Numerical interpolation and chemical potentials}

The first-principles calculations provide the DFT formation enthalpies
only at discrete hydrogen concentrations. Direct differentiation of these
discrete data would introduce numerical noise into
$\mu_{\mathrm H}^{\alpha}$, which is particularly problematic because the
equilibrium pressure depends exponentially on the hydrogen chemical
potential. Therefore, the concentration-dependent DFT enthalpies entering
the free-energy expression were interpolated before evaluating the
thermodynamic derivatives. Shape-preserving piecewise cubic
Hermite interpolating polynomial (PCHIP)
\cite{FritschCarlson1980} was employed
to represent the DFT formation enthalpy as a continuous function of hydrogen
concentration,

\begin{equation}
H_{\mathrm{DFT}}^{\alpha}
=
H_{\mathrm{DFT}}^{\alpha}(c_{\mathrm H}).
\label{eq:H_DFT_interpolated}
\end{equation}

PCHIP was selected because it preserves the local shape and monotonicity of
the underlying data and avoids the overshooting that can occur with
conventional cubic splines. Additionally, this procedure provides 
a consistent continuous thermodynamic
representation while avoiding the artificial
oscillations that may arise from higher-order
polynomial interpolation. This property is particularly important for the
present application because small numerical oscillations in
$G^{\alpha}(c_{\mathrm H},T)$ can be amplified by
differentiation and subsequently produce substantial variations in the
calculated equilibrium pressure.
The interpolated DFT enthalpies were evaluated on a dense
hydrogen-concentration grid for each alloy composition and crystallographic
phase. The free-energy function was then constructed using
Eq.~\eqref{eq:M0_free_energy}, with the temperature-dependent hydrogen-gas
contributions and the phase-dependent configurational entropy evaluated
consistently on the same concentration grid. A mild Gaussian smoothing
operation was applied to the resulting free-energy curves before numerical
differentiation in order to suppress residual numerical noise while
preserving their overall thermodynamic shape. The smoothing was performed by convolution with a Gaussian kernel. 



\subsection{\label{sec:common_tangent}Common-tangent construction}

Consider two phases, $\alpha$ and $\beta$, with hydrogen concentrations
$c_{\mathrm H}^{\alpha}$ and $c_{\mathrm H}^{\beta}$, respectively. The
slope of the line connecting the corresponding points on the two
Gibbs free-energy curves is

\begin{equation}
m
=
\frac{
G^{\beta}(c_{\mathrm H}^{\beta},T)
-
G^{\alpha}(c_{\mathrm H}^{\alpha},T)
}{
c_{\mathrm H}^{\beta}
-
c_{\mathrm H}^{\alpha}
}.
\label{eq:common_tangent}
\end{equation}

At phase coexistence, this line is a common tangent to the two free-energy
curves. Consequently, its slope must simultaneously equal the hydrogen
chemical potential of both phases,

\begin{equation}
m
=
\mu_{\mathrm H}^{\alpha}
=
\mu_{\mathrm H}^{\beta}.
\label{eq:tangent_condition}
\end{equation}

However, a
common tangent requires not only equal slopes, corresponding to
equal hydrogen chemical potentials, but also equal intercepts,
corresponding to equal metal chemical potentials. 
Consequently, the common-tangent construction simultaneously
determines the equilibrium coexistence compositions and the
chemical potentials of both components. In the present
implementation, this construction was performed independently
for the BCC--BCT and BCT--FCC phase pairs. Here, BCT denotes the
tetragonal phase with tetrahedral hydrogen occupation, which
participates in the sequence of phase equilibria considered in
this work.

Rather than relying on an iterative solution of the common-tangent
equations, all possible pairs of compositions from the two
phase-specific free-energy grids were evaluated simultaneously.
For each pair, the slope of the connecting line was calculated
according to Eq.~\eqref{eq:common_tangent}. Since this line passes
through both selected points, its intercept is necessarily the
same for the two phases. Thus, the common-tangent condition is
satisfied when the slope of the connecting line also matches
the hydrogen chemical potential of each phase. The deviation
from this condition was quantified through

\begin{equation}
\varepsilon_{\mathrm{tan}}
=
\left|
\mu_{\mathrm H}^{\alpha}-m
\right|
+
\left|
\mu_{\mathrm H}^{\beta}-m
\right|.
\label{eq:tangent_error}
\end{equation}

The pair yielding the minimum $\varepsilon_{\mathrm{tan}}$ was
selected as the numerical coexistence condition. This fully
vectorized procedure avoids the convergence difficulties
associated with iterative nonlinear searches and allows the
coexistence compositions to be obtained directly from the
calculated free-energy curves.




For the first plateau, the coexistence compositions are denoted by
$c_{\mathrm H}^{\mathrm{BCC}}$ and
$c_{\mathrm H}^{\mathrm{BCT}}$. For the second plateau, they are denoted
by $c_{\mathrm H}^{\mathrm{BCT}}$ and
$c_{\mathrm H}^{\mathrm{FCC}}$. The width of the second plateau is
therefore

\begin{equation}
\Delta c_{\mathrm H}^{(2)}
=
c_{\mathrm H}^{\mathrm{FCC}}
-
c_{\mathrm H}^{\mathrm{BCT}}.
\label{eq:plateau_width}
\end{equation}

The same construction consequently provides not only the equilibrium
pressure but also the composition interval over which the two phases
coexist. 

\subsection{Equilibrium pressure}




From equations~(\ref{eq:muH_gas}) and  (\ref{eq:muH_equilibrium}) and using the common hydrogen chemical potential obtained from the tangent
construction therefore gives

\begin{equation}
\ln
\left(
\frac{P_{\mathrm{eq}}}{P^0}
\right)
=
\frac{2\mu_{\mathrm H}}{RT},
\label{eq:P_eq}
\end{equation}
and hence

\begin{equation}
P_{\mathrm{eq}}
=
P^0
\exp
\left(
\frac{2\mu_{\mathrm H}}{RT}
\right).
\label{eq:P_eq_exp}
\end{equation}

In the implementation, the reference pressure is taken consistently with
the pressure convention used for the experimental PCT data. Equivalently,
the plateau pressure can be obtained directly from the slope of the common
tangent,

\begin{equation}
P_{\mathrm{eq}}
=
P^0
\exp
\left[
\frac{2}{RT}
\frac{
G^{\beta}(c_{\mathrm H}^{\beta},T)
-
G^{\alpha}(c_{\mathrm H}^{\alpha},T)
}{
c_{\mathrm H}^{\beta}
-
c_{\mathrm H}^{\alpha}
}
\right].
\label{eq:P_plateau_tangent}
\end{equation}

This expression makes explicit the connection between the DFT-based
thermodynamic description and the macroscopic PCT response. The
DFT-derived formation enthalpies, together with the hydrogen-gas
contributions and configurational entropy included in the thermodynamic model, determine the
shape of the phase-specific free-energy curves. The common tangent
determines the coexistence compositions and the common hydrogen chemical
potential, which is then converted into the equilibrium hydrogen pressure
through Eq.~\eqref{eq:P_eq}.





\section{Results and Discussion}

\subsection{Calibration of the thermodynamic  model}

The thermodynamic model contains a single adjustable parameter,
$A_{\mathrm{298K}}$, which acts as a global correction to the hydrogen
reference contribution in the Gibbs free-energy expression. This
parameter is expressed in $\mathrm{kJ\,mol^{-1}_{H_2}}$ and enters
Eq.~\eqref{eq:M0_free_energy} through the term proportional to
$c_{\mathrm H}/2$. 

A global optimization was performed using the complete experimental
dataset, comprising seven alloy compositions and 22
alloy--temperature conditions. All 22 experimental alloy--temperature conditions listed in
Table~\ref{tab:optimization_dataset} were included in the global
optimization of $A_{\mathrm{298K}}$. No alloy composition or
experimental condition was excluded to construct an independent
test set. The optimized parameter therefore represents the best
global fit to the complete experimental dataset considered in this
work. For each trial value of $A_{\mathrm{298K}}$,
the Gibbs free-energy curves were reconstructed, and the BCT--FCC
common-tangent construction was applied to determine the second-plateau
equilibrium pressure and hydrogen-composition width, $
\Delta c_{\mathrm H}^{(2)}
=
c_{\mathrm H}^{\mathrm{FCC}}
-
c_{\mathrm H}^{\mathrm{BCT}}$.
\begin{table}[ht]
\centering
\caption{Alloy compositions and experimental temperatures included
in the global optimization of the thermodynamic model.}
\label{tab:optimization_dataset}
\begin{tabular}{lc}
\hline
Alloy & Temperature (K) \\
\hline
$\mathrm{TiVNb}_{80}\mathrm{CrAl}_{20}$\cite{Ferreira2023TiVNbCrAl} & 263, 273, 298 \\
$\mathrm{TiVNb}_{90}\mathrm{CrAl}_{10}$\cite{Ferreira2023TiVNbCrAl} & 373, 393, 408, 423 \\
$\mathrm{TiVNb}_{85}\mathrm{Cr}_{15}$\cite{Silva2021TiVNbVEC} & 373, 423, 473 \\
$\mathrm{TiVNb}_{75}\mathrm{Cr}_{25}$\cite{Nygard2021TiVCrNb} & 348, 373, 388 \\
$\mathrm{TiVNb}_{70}\mathrm{Cr}_{30}$\cite{Strozi2023TiVNbCr} & 303, 323, 343 \\
$\mathrm{TiVNb}_{67}\mathrm{Cr}_{33}$\cite{Silva2023RoomTemperatureReversibility} & 298, 323, 348 \\
$\mathrm{TiVNb}_{65}\mathrm{Cr}_{35}$\cite{Strozi2023TiVNbCr} & 303, 323, 343 \\
\hline
\end{tabular}
\end{table}

The pressure and width were optimized simultaneously using the
objective function

\begin{equation}
\begin{aligned}
\mathcal{L}(A_{\mathrm{298K}})
={}&
\sum_{s,T}
\left[
\ln(P_{\mathrm{eq}}^{\mathrm{model}})
-
\ln(P_{\mathrm{eq}}^{\mathrm{exp}})
\right]^2
\\
&+
\sum_{s,T}
\left[
\frac{
\Delta c_{\mathrm H}^{(2),\mathrm{model}}
-
\Delta c_{\mathrm H}^{(2),\mathrm{exp}}
}{
\Delta c_{\mathrm H}^{(2),\mathrm{exp}}
}
\right]^2.
\end{aligned}
\label{eq:objective_M0}
\end{equation}

Here, $s$ denotes the alloy composition and $T$ the experimental
temperature. The logarithmic pressure residual accounts for the
several-orders-of-magnitude variation in equilibrium pressures,
emphasizing relative rather than absolute deviations. The width
residual is normalized by the corresponding experimental value.
Both contributions are dimensionless, and no additional weighting
factor was introduced.
The optimization was performed using the differential-evolution
algorithm \cite{StornPrice1997}, with the search interval

\begin{equation}
-60 \leq A_{\mathrm{298K}} \leq 60
\quad
\mathrm{kJ\,mol^{-1}_{H_2}}.
\label{eq:Afit_bounds}
\end{equation}

For each trial value, the free-energy curves and their composition
derivatives were evaluated, followed by the BCT--FCC common-tangent
construction. Trial values producing nonphysical results, including
non-finite or non-positive equilibrium pressures or plateau widths,
were penalized through the objective function.
The global optimization yielded

\begin{equation}
A_{\mathrm{298K}}
=
33.55294\ \mathrm{kJ\,mol^{-1}_{H_2}}.
\label{eq:Afit_opt}
\end{equation}

The optimized value lies within the prescribed search interval and
does not coincide with either bound. The same parameter was used for
all seven compositions and 22 experimental conditions. 
Simultaneous optimization of the equilibrium pressure and plateau
width constrains distinct features of the free-energy landscape.
While the equilibrium pressure is governed by the common hydrogen
chemical potential, the plateau width is determined by the separation
between the two coexistence compositions. Therefore, considering both
observables provides a more restrictive constraint than
pressure-only optimization. The calculated equilibrium pressures, plateau widths, and complete
PCT curves are compared with the corresponding experimental results
in the following sections. These comparisons assess the overall
agreement achieved by the thermodynamic model  formulation across the investigated
compositions and temperatures.

\subsection{Thermodynamic analysis of TiVNb75Cr25 at 373 K}

Figure~\ref{fig:thermodynamic_diagnostic_TiVNb75Cr25_373K}
provides a thermodynamic analysis of the $\mathrm{TiVNb}_{75}\mathrm{Cr}_{25}$
alloy at $373$~K using our thermodynamic model. The three panels summarize
the main steps connecting the first-principles energetic description
to the phase-equilibrium conditions: the DFT formation enthalpies,
the configurational entropy, and the resulting Gibbs free-energy
landscape. Together, these quantities determine the relative
stability of the BCC, BCT, and FCC configurations and the
coexistence compositions obtained from the common-tangent
construction.

\subsubsection{DFT formation enthalpies}

\begin{figure}[H]
    \centering
    \includegraphics[
        width=0.88\textwidth
    ]{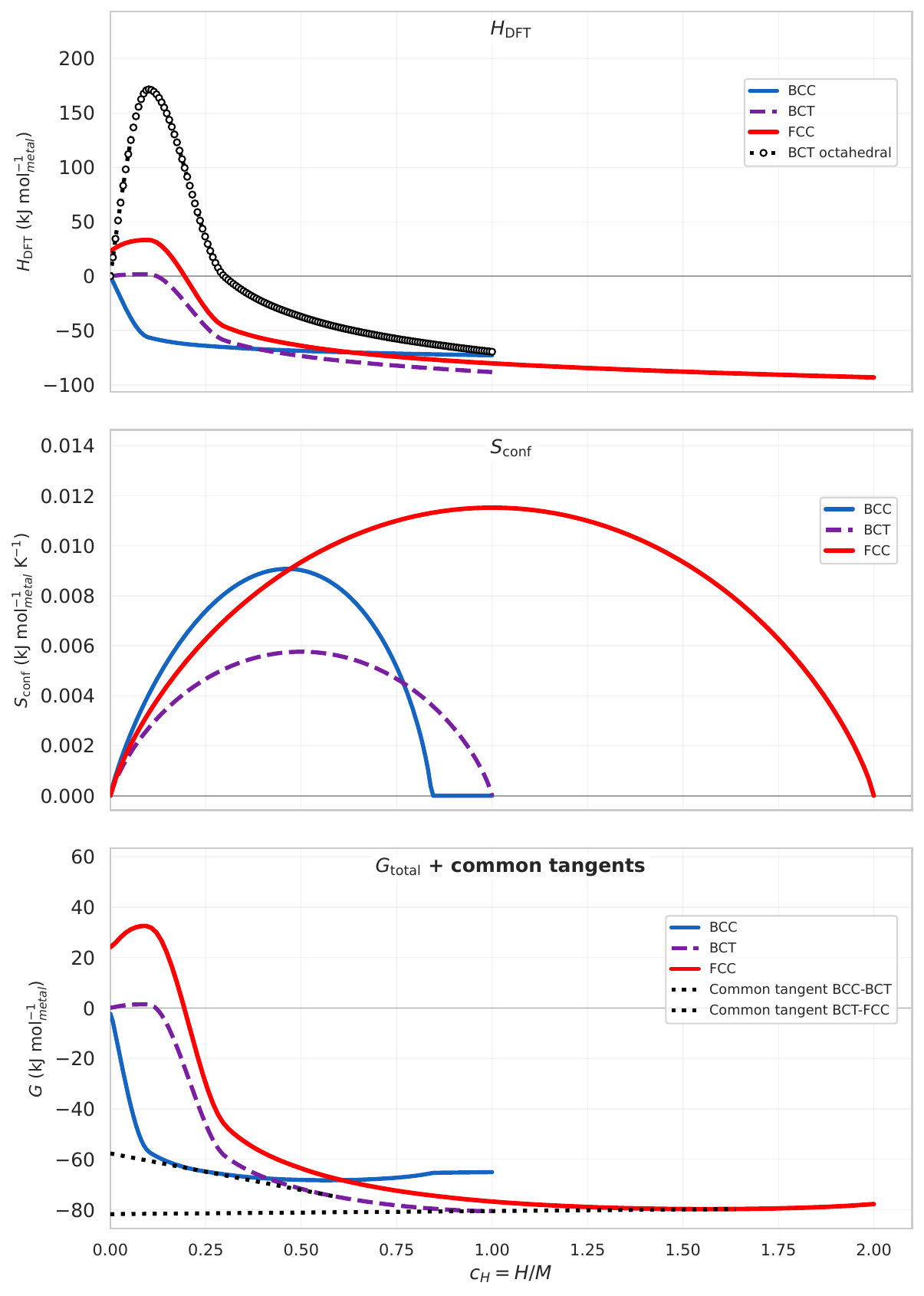}
        \caption{
    Thermodynamic description of the $\mathrm{TiVNb}_{75}\mathrm{Cr}_{25}$ alloy at $T=373$~K using our proposed free-energy model.
    (a) DFT formation enthalpies, $H_{\mathrm{DFT}}$, for the BCC, BCT tetrahedral, and FCC structures. The BCT octahedral configuration is shown for comparison but is not considered a competing equilibrium phase.
    (b) Configurational entropy, $S_{\mathrm{conf}}$, as a function of hydrogen concentration for the three competing phases.
    (c) Total Gibbs free energy, $G_{\mathrm{total}}$, including the common tangents associated with the BCC--BCT and BCT--FCC phase equilibria. The tangent lines are extended to $c_{\mathrm H}=0$ for visualization.
    All energies are expressed per mole of metal atoms.
    }
    \label{fig:thermodynamic_diagnostic_TiVNb75Cr25_373K}
\end{figure}

Figure~\ref{fig:thermodynamic_diagnostic_TiVNb75Cr25_373K}~(a) shows the DFT formation enthalpies,
$H_{\mathrm{DFT}}^\alpha(c_{\mathrm H})$, for the BCC, BCT
tetrahedral, and FCC configurations. The three branches exhibit
distinct concentration-dependent energetic profiles, reflecting
the different local environments and interstitial-site arrangements
associated with hydrogen incorporation in each structure.
The BCC branch is energetically favorable at low hydrogen
concentrations, whereas the BCT tetrahedral and FCC branches
become increasingly competitive as the hydrogen content increases.
The FCC curve starts at a positive enthalpy at $c_{\mathrm H}=0$,
in contrast to the BCC reference, whose enthalpy is zero at this
composition. This difference reflects the energy required to
transform the hydrogen-free alloy from its most stable BCC
configuration to the FCC structure. Accordingly, the energy
difference between the hydrogen-free metallic structures is
retained as a constant offset in the FCC formation-enthalpy curve,
ensuring that all phase branches are expressed on a common
energetic reference.
The BCT octahedral configuration is also displayed for comparison.
Its formation enthalpy remains substantially higher than that of
the BCT tetrahedral configuration over the investigated
concentration range, particularly at low hydrogen contents.
Therefore, it is retained as an alternative DFT structural
configuration but is not treated as an independent competing
phase in the phase-equilibrium calculations. Only the BCC,
BCT tetrahedral, and FCC branches enter the common-tangent
construction.

\subsubsection{Configurational entropy}

Figure~\ref{fig:thermodynamic_diagnostic_TiVNb75Cr25_373K}~(b) presents the configurational entropy,
$S_{\mathrm{conf}}^\alpha(c_{\mathrm H})$, associated with
hydrogen occupation of interstitial sites. The concentration
dependence differs among the three phases because the number
and accessibility of interstitial sites depend on the
crystallographic structure.
The BCC branch exhibits a maximum at an intermediate hydrogen
concentration and decreases to zero as the accessible interstitial
sites become fully occupied. This behavior reflects the
concentration-dependent site-blocking treatment adopted for
the BCC structure. The BCT and FCC branches display distinct
profiles according to their respective interstitial-site
populations and configurational models. In particular, the
FCC branch extends to higher hydrogen concentrations and
exhibits a broader configurational-entropy contribution.
These differences show that hydrogen configurational entropy
is phase dependent and therefore modifies the relative stability
of the structural branches at finite temperature. The entropy
contribution is especially relevant when comparing phases whose
enthalpic energies are close, since it can change their relative
free-energy positions through the term $-T S_{\mathrm{conf}}^\alpha$.

\subsubsection{Gibbs free energy and common-tangent constructions}

Figure~\ref{fig:thermodynamic_diagnostic_TiVNb75Cr25_373K}~(c) shows the resulting Gibbs free-energy curves,
$G_{\mathrm{total}}^\alpha(c_{\mathrm H})$, at $373$~K,
together with the common tangents associated with the
BCC--BCT and BCT--FCC equilibria. The free-energy landscape
combines the DFT formation enthalpies with the common
finite-temperature molecular hydrogen reference and the
phase-dependent configurational entropy contribution.
The relative positions of the free-energy branches vary with
hydrogen concentration, allowing the BCC, BCT, and FCC
configurations to participate successively in the equilibrium
description. The phase boundaries are determined by the
common-tangent construction rather than by the direct
intersection of the free-energy curves. The first common
tangent connects the BCC and BCT branches, defining the
coexistence compositions associated with the first plateau.
The second common tangent connects the BCT and FCC branches
and determines the coexistence compositions of the second
plateau.
For the BCT--FCC equilibrium, the calculated plateau width is
given by
$
\Delta c_{\mathrm H,\mathrm{calc}}
=
c_{\mathrm H}^{\mathrm{FCC}}
-
c_{\mathrm H}^{\mathrm{BCT}}$, 
where the two concentrations correspond to the contact points
of the common tangent with the BCT and FCC free-energy
branches, respectively. For $\mathrm{TiVNb}_{75}\mathrm{Cr}_{25}$
at $373$~K, the calculated equilibrium pressure of this
transition is $P_{\mathrm{calc}}=2.210$~bar.
The tangent lines are extended to $c_{\mathrm H}=0$ in
Figure~\ref{fig:thermodynamic_diagnostic_TiVNb75Cr25_373K}~(c) to facilitate visualization of their slopes and
intercepts. This graphical extension does not change the
coexistence compositions or the thermodynamic equilibrium
conditions, which are determined by the tangent contact
points with the corresponding free-energy branches.

Overall, the three panels illustrate how the phase-dependent
DFT energetics and configurational entropy combine to produce
the Gibbs free-energy landscape from which the phase
coexistence compositions and equilibrium pressures are
obtained. The detailed decomposition of the molecular
hydrogen reference, entropy contributions, hydrogen chemical
potential, and pressure profile is provided in the
Supplementary Material.

\subsection{Representative PCT diagrams}

\begin{figure}[H]
    \centering
    \includegraphics[width=\linewidth]{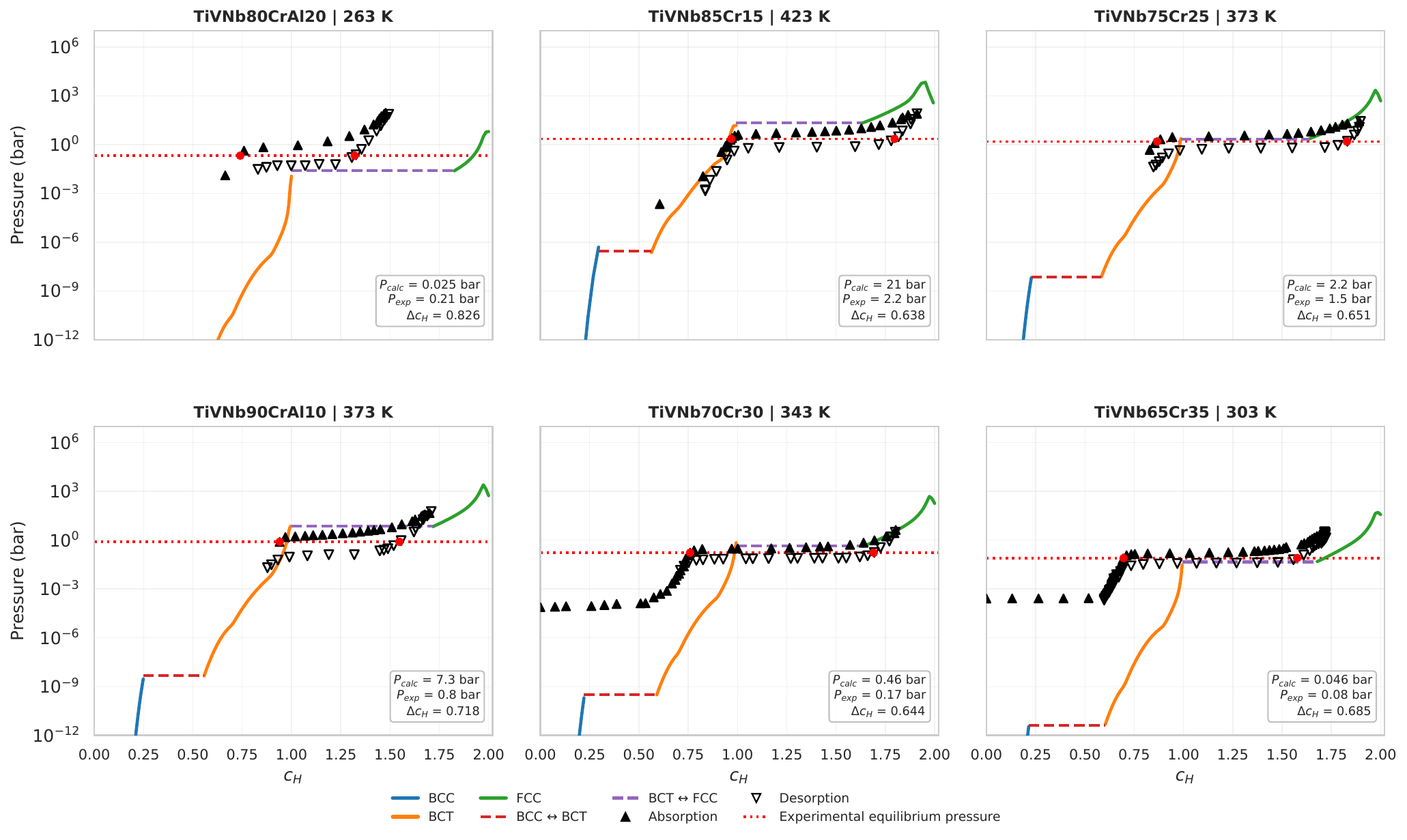}
    \caption{Representative pressure-composition-temperature (PCT)
    diagrams calculated using our  thermodynamic model for six
    Ti--V--Nb-based multicomponent alloys at the indicated temperatures:
    (a) $\mathrm{TiVNb}_{80}\mathrm{CrAl}_{20}$ at 263 K
    \cite{Ferreira2023TiVNbCrAl};
    (b) $\mathrm{TiVNb}_{85}\mathrm{Cr}_{15}$ at 423 K
    \cite{Silva2021TiVNbVEC};
    (c) $\mathrm{TiVNb}_{75}\mathrm{Cr}_{25}$ at 373 K
    \cite{Nygard2021TiVCrNb};
    (d) $\mathrm{TiVNb}_{90}\mathrm{CrAl}_{10}$ at 373 K
    \cite{Ferreira2023TiVNbCrAl};
    (e) $\mathrm{TiVNb}_{65}\mathrm{Cr}_{35}$ at 303 K
    \cite{Strozi2023TiVNbCr};
    and (f) $\mathrm{TiVNb}_{70}\mathrm{Cr}_{30}$ at 343 K
    \cite{Strozi2023TiVNbCr}.
    Solid curves represent the calculated pressure-composition
    branches associated with the BCC, BCT, and FCC phases.
    Dashed horizontal lines indicate the calculated equilibrium
    pressures obtained from common-tangent constructions for the
    BCC--BCT and BCT--FCC transformations. Black upward-pointing
    filled and downward-pointing open triangles represent
    experimental absorption and desorption data, respectively.
    The red dotted horizontal line and the corresponding
    $P_{\mathrm{exp}}$ label indicate the experimental equilibrium
    pressure of the second plateau, which was calculated by taking the average between absorption and desorption data. The $P_{\mathrm{calc}}$ label
    denotes the calculated equilibrium pressure of the BCT--FCC
    plateau, while $\Delta c_{\mathrm{H}}$ denotes its calculated
    width, defined as the difference between the hydrogen
    concentrations at the two common-tangent contact points.}
    \label{fig:M0_representative_PCT}
\end{figure}

Figure~\ref{fig:M0_representative_PCT} presents six representative pressure–composition–temperature (PCT) diagrams obtained with the thermodynamic model for selected Ti–V–Nb-based multicomponent alloys. The systems and temperatures were chosen to illustrate the model predictions over a range of compositions and experimental conditions, while the results for the remaining 16 systems/conditions are provided in the Supplementary Material.  The calculated phase sequence is BCC \(\rightleftharpoons\) BCT \(\rightleftharpoons\) FCC, with the two equilibrium plateaus determined from common-tangent constructions between the corresponding free-energy curves.
The first equilibrium plateau is associated with the BCC–BCT transformation, whereas the second corresponds to the BCT–FCC transformation. The calculated pressure of the second plateau, \(P_{\mathrm{calc}}\), is therefore obtained from the common tangent connecting the BCT and FCC free-energy branches. The corresponding experimental equilibrium pressure, \(P_{\mathrm{exp}}\), is indicated by the red dotted horizontal line in each panel. Thus, the comparison between \(P_{\mathrm{calc}}\) and \(P_{\mathrm{exp}}\) directly assesses the model's ability to reproduce the experimental equilibrium pressure of the second plateau.
The calculated width of the second plateau, $\Delta c_{H}$, is also reported in each panel. This quantity is defined as the difference between the hydrogen concentrations at the two contact points of the common tangent connecting the BCT and FCC free-energy curves. It therefore represents the calculated composition interval associated with the BCT–FCC equilibrium, rather than a width estimated from the experimental PCT data.

For TiVNb$_{80}$CrAl$_{20}$ at 263 K, the calculated second-plateau pressure is 0.025 bar, whereas the experimental value is 0.21 bar. The corresponding calculated plateau width is $\Delta(H/M) = 0.826$. For TiVNb$_{65}$Cr$_{35}$ at 303 K, the calculated and experimental second-plateau pressures are 0.046 and 0.08 bar, respectively, and the calculated plateau width is 0.685. A larger discrepancy is observed for TiVNb$_{85}$Cr$_{15}$ at 423 K, for which the model predicts a second-plateau pressure of 21 bar, compared with the experimental value of 2.2 bar; the calculated plateau width is 0.638.
The other three panels of Figure~\ref{fig:M0_representative_PCT} show the results for TiVNb$_{90}$CrAl$_{10}$ at 373 K, TiVNb$_{75}$Cr$_{25}$ at 373 K, and TiVNb$_{70}$Cr$_{30}$ at 343 K. For these alloys, the calculated second-plateau pressures are 7.3, 2.2, and 0.46 bar, respectively, while the corresponding experimental values are 0.8, 1.5, and 0.17 bar. The calculated plateau widths are 0.718, 0.651, and 0.644, respectively.

Overall, the selected diagrams illustrate that the thermodynamic model captures the principal thermodynamic features associated with the two-step hydrogenation sequence and provides calculated second-plateau widths across the selected alloy systems. The agreement between calculated and experimental second-plateau pressures, however, varies with alloy composition and temperature. These differences motivate the global assessment of pressure and plateau-width predictions presented in the following sections. 
\subsection{Global assessment of the second-plateau equilibrium pressure}

\begin{figure}[t]
    \centering
    \includegraphics[width=\linewidth]
    {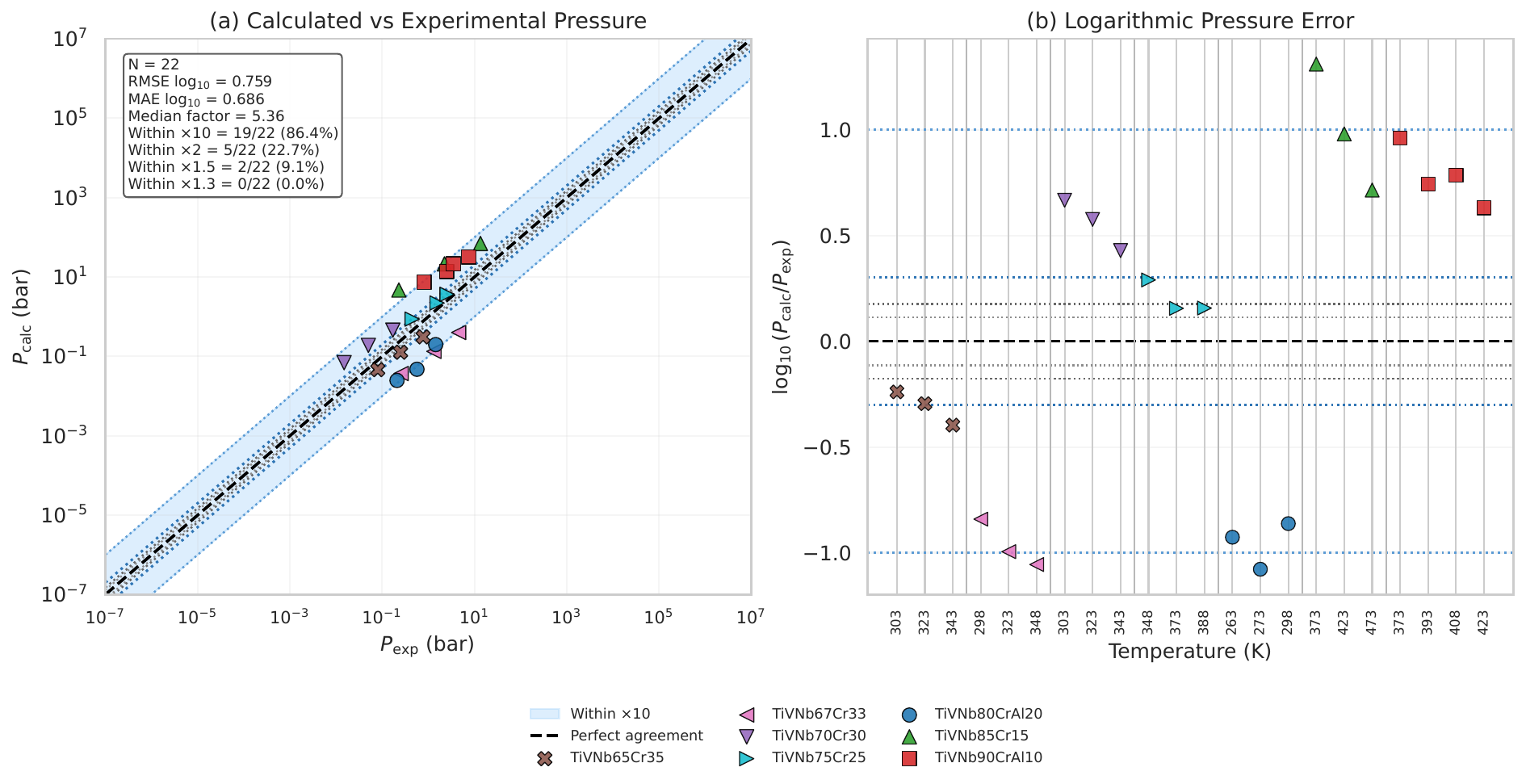}
    \caption{
    Global assessment of the thermodynamic model prediction of the second-plateau
    equilibrium pressure over the 22 experimental PCT conditions
    corresponding to seven alloy compositions.
    (a) Calculated versus experimental equilibrium pressures
    on logarithmic axes. The dashed diagonal indicates perfect
    agreement, while the shaded regions indicate deviations
    within factors of 10, 2, and 1.5.
    (b) Logarithmic pressure error,
    $\log_{10}(P_{\mathrm{calc}}/P_{\mathrm{exp}})$,
    as a function of temperature for the seven alloy compositions.
    The model gives
    $\mathrm{RMSE}_{\log_{10}}=0.759$,
    $\mathrm{MAE}_{\log_{10}}=0.686$, and a median multiplicative
    deviation of 5.36. A total of 19/22 conditions (86.4\%)
    are within a factor of 10, 5/22 (22.7\%) are within a factor
    of 2, and 2/22 (9.1\%) are within a factor of 1.5.
    }
    \label{fig:M0_global_pressure_assessment}
\end{figure}

The predictive performance of the thermodynamic model was assessed over the
complete set of 22 experimental PCT conditions corresponding to
seven alloy compositions. Figure~\ref{fig:M0_global_pressure_assessment}
compares the calculated and experimental equilibrium pressures of
the second plateau over the investigated composition and temperature
ranges. Since the single parameter $A_{\mathrm{298K}}$ was calibrated
using this complete dataset, the results represent a global
assessment of the model rather than an independent validation.

Figure~\ref{fig:M0_global_pressure_assessment}(a) shows the
calculated versus experimental pressures on logarithmic axes.
The calculated values reproduce the broad experimental pressure
range and follow the overall diagonal trend, indicating that the
thermodynamic model formulation captures the general pressure scale across the
investigated conditions. However, a substantial dispersion around
the line of perfect agreement remains. The root-mean-square error
on the base-10 logarithmic scale is
$\mathrm{RMSE}_{\log_{10}}=0.759$, while the corresponding
mean absolute error is
$\mathrm{MAE}_{\log_{10}}=0.686$.
The median multiplicative deviation between calculated and
experimental pressures is a factor of 5.36.

Considering the multiplicative agreement criteria, 19 of the
22 conditions (86.4\%) are reproduced within a factor of 10.
However, only 5 conditions (22.7\%) fall within a factor of 2,
and 2 conditions (9.1\%) are within a factor of 1.5.
None of the calculated pressures falls within a factor of 1.3
of the corresponding experimental value. These results indicate
that, although the model captures the overall pressure range,
its quantitative accuracy varies considerably among the
investigated alloy compositions and temperatures.

The composition- and temperature-resolved logarithmic residuals
shown in panel~(b),
\begin{equation}
\varepsilon_P
=
\log_{10}
\left(
\frac{P_{\mathrm{calc}}}{P_{\mathrm{exp}}}
\right),
\end{equation}
provide further insight into the origin of these deviations.
The residuals are not randomly distributed around zero.
In particular, the $\mathrm{TiVNb}_{80}\mathrm{CrAl}_{20}$
conditions are predominantly associated with negative pressure
errors, indicating systematic underestimation of the experimental
pressures. In contrast, several conditions corresponding to
$\mathrm{TiVNb}_{85}\mathrm{Cr}_{15}$ and
$\mathrm{TiVNb}_{90}\mathrm{CrAl}_{10}$ exhibit positive
deviations, indicating overestimation. The remaining compositions
show intermediate residual patterns.

The remaining deviations in the equilibrium pressures should be considered in light of the uncertainties and approximations inherent to both the experimental measurements and the theoretical framework. On the experimental side, differences between nominal and actual alloy compositions may arise from segregation, incomplete homogenization, synthesis kinetics, diffusion, and characterization uncertainties. On the theoretical side, the CPA--KKR calculations do not explicitly account for atomic relaxation or hydrogen-induced lattice distortions, while vibrational and anharmonic contributions are also neglected. Furthermore, since the equilibrium pressure depends exponentially on the hydrogen chemical potential, relatively small differences in this quantity may lead to substantial deviations in the predicted pressures. Therefore, the observed discrepancies likely reflect the combined influence of these factors rather than a single source of error. Despite these limitations, the model captures the broad pressure range and the general variation of the second plateau across the investigated conditions.


\subsection{Global assessment of the second-plateau hydrogen-composition width}

\begin{figure}[t]
    \centering
    \includegraphics[width=\linewidth]
    {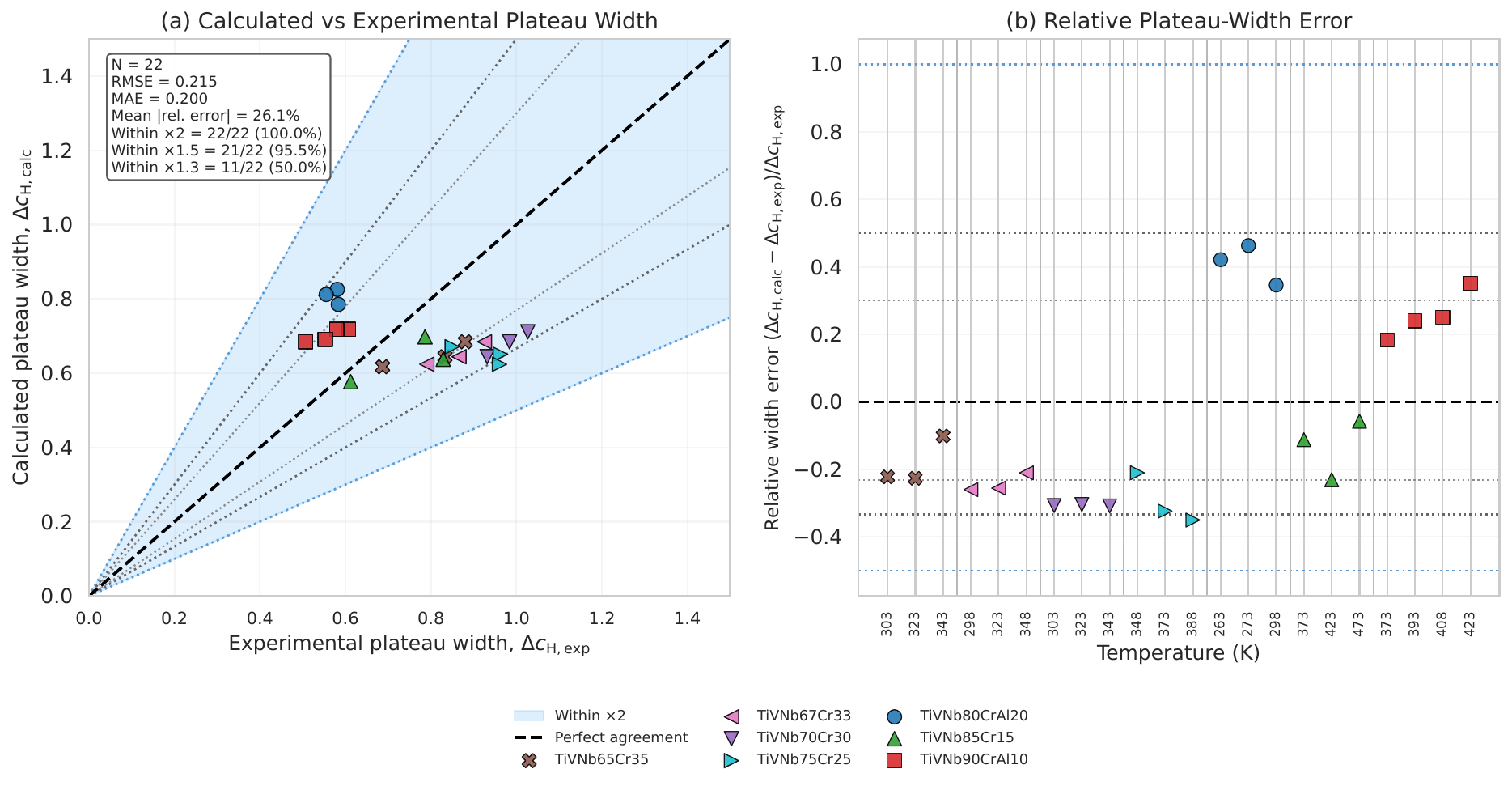}
    \caption{
    Global assessment of the thermodynamic model prediction of the second-plateau
    hydrogen-composition width over the 22 experimental PCT
    conditions corresponding to seven alloy compositions.
    (a) Calculated versus experimentally estimated plateau
    widths. The dashed diagonal indicates perfect agreement,
    and the shaded regions denote deviations within factors
    of 2, 1.5, and 1.3.
    (b) Relative plateau-width error,
    $(\Delta c_{H,\mathrm{calc}}-\Delta c_{H,\mathrm{exp}})
    /\Delta c_{H,\mathrm{exp}}$, as a function of temperature
    for the seven alloy compositions.
    The experimental width is estimated from the intersections
    of the horizontal line at the experimental second-plateau
    equilibrium pressure with the experimental absorption
    and/or desorption data.
    The model gives $\mathrm{RMSE}=0.215$,
    $\mathrm{MAE}=0.200$, and a mean absolute relative error
    of 26.1\%. All 22 conditions are within a factor of 2,
    21/22 (95.5\%) are within a factor of 1.5, and
    11/22 (50.0\%) are within a factor of 1.3.
    }
    \label{fig:M0_global_width_assessment}
\end{figure}

The calculated second-plateau hydrogen-composition widths are
compared with the corresponding experimental estimates in
Fig.~\ref{fig:M0_global_width_assessment}. In general, the thermodynamic model reproduces the plateau widths
with substantially small deviations across the complete
22-condition dataset.
The experimental plateau width,
$\Delta c_{H,\mathrm{exp}}$, is estimated directly from the
experimental PCT data using the horizontal line corresponding
to the experimental second-plateau equilibrium pressure,
$P_{\mathrm{exp}}$. The two hydrogen concentrations defining
this width are identified at the intersections of this
horizontal line with the experimental data points. These
intersections may occur on either the absorption or the
desorption branch; no specific branch is assigned a priori
to either limit of the plateau. The experimental width is
then estimated as the difference between the higher and lower
hydrogen concentrations at these intersections:

\begin{equation}
\Delta c_{H,\mathrm{exp}}
=
c_{H,\mathrm{exp}}^{\mathrm{high}}
-
c_{H,\mathrm{exp}}^{\mathrm{low}}.
\end{equation}

This procedure provides an approximate experimental estimate
of the second-plateau composition interval, since the limits
are determined from the intersections of the experimental
pressure line with the measured absorption and/or desorption
data. It should therefore be distinguished from the calculated
plateau width, which is obtained thermodynamically from the
two BCT--FCC common-tangent contact compositions.

Figure~\ref{fig:M0_global_width_assessment}(a) compares the
calculated and experimentally estimated widths, with the
dashed diagonal representing perfect agreement. The
root-mean-square error is $\mathrm{RMSE}=0.215$, and the
mean absolute error is $\mathrm{MAE}=0.200$. The mean
absolute relative error is 26.1\%. All 22 conditions are
reproduced within a factor of 2, while 21 conditions (95.5\%)
fall within a factor of 1.5. Furthermore, 11 conditions
(50.0\%) are within a factor of 1.3 of their experimental
estimates.

The relative width errors shown in panel~(b) are defined as
\begin{equation}
\varepsilon_{\Delta c_H}
=
\frac{
\Delta c_{H,\mathrm{calc}}
-
\Delta c_{H,\mathrm{exp}}
}{
\Delta c_{H,\mathrm{exp}}
}.
\end{equation}
Although the magnitude of these deviations is considerably
smaller than that observed for the equilibrium pressure,
the residuals also exhibit composition-dependent patterns.
The $\mathrm{TiVNb}_{80}\mathrm{CrAl}_{20}$ and
$\mathrm{TiVNb}_{90}\mathrm{CrAl}_{10}$ conditions are
predominantly associated with positive width errors,
indicating overestimation of the experimental plateau
widths. In contrast, several of the higher-Cr compositions
exhibit predominantly negative deviations. Thus, the width
errors are not entirely random, even though their overall
magnitude is relatively small.

The comparatively good reproduction of the plateau width
is particularly relevant because this quantity is not
independently fitted at each temperature. Instead, the
calculated width emerges directly from the two hydrogen
concentrations defining the BCT--FCC common-tangent contacts.
Its agreement with the experimentally estimated values
therefore indicates that the free-energy landscape
captures an important part of the concentration dependence
governing the BCT--FCC coexistence interval across different
alloy compositions and temperatures.


\section{Comparative Assessment of Thermodynamic Models}

To further contextualize the performance of the present model, we compare it with three representative approaches from the literature: the random site occupancy (RSO) model proposed by Zepon et al.~\cite{Zepon2021}, the discrete site energy (DSE) model developed by Pedroso et al.~\cite{PedrosoDSE}, and the automated DFT--CALPHAD framework introduced by Hannappel et al.~\cite{Hannappel2025CALPHAD,Hannappel2026}

The RSO model adopts a mean-field approximation in which all interstitial sites are treated as a single average site whose properties are determined by the overall alloy composition. Although computationally inexpensive, its assumption of a constant partial molar enthalpy of hydrogen results in a linear dependence of the mixing enthalpy 
on hydrogen concentration. This simplification can substantially overestimate the two-phase plateau width, $\Delta c_{\mathrm H}$, in PCT diagrams. To account for local chemical heterogeneity, the DSE model explicitly evaluates the hydrogen interaction energies, associated with the 35 distinct tetrahedral environments in quaternary BCC alloys using DFT supercell calculations. By minimizing the Gibbs free energy under para-equilibrium conditions, this approach captures the preferential occupation of low-energy interstitial sites, particularly those surrounded by hydride-forming elements, reducing the overestimation of the plateau width by approximately 15--20\% relative to the RSO model. However, this improved local resolution requires 16-atom supercells and the evaluation of numerous configurations for each crystalline phase ($\alpha$, $\beta$, and $\delta$), leading to a substantial computational time. Moreover, the number of distinct local environments increases combinatorially as additional alloying elements are introduced, making extensions to systems containing five or more components increasingly demanding.

A different strategy was introduced by Hannappel et al.~\cite{Hannappel2025CALPHAD,Hannappel2026}, who developed an automated framework combining DFT calculations with the Compound Energy Formalism (CEF) within the CALPHAD approach to predict hydrogen para-equilibrium thermodynamics in multicomponent systems. This framework reduces the manual effort associated with conventional CALPHAD assessments and enables the calculation of pressure--composition isotherms. However, the accuracy of multicomponent predictions remains dependent on the availability and transferability of thermodynamic parameters, particularly when higher-order interactions in disordered solid solutions are not explicitly described.

The present model addresses several of these computational and thermodynamic challenges through KKR--CPA method. By describing substitutional chemical disorder in the metallic sublattice through an effective medium, the method enables the direct calculation of multicomponent alloy energetics using a primitive unit cell, independently of the number of alloying elements. This avoids the need for explicit configurational sampling through large supercells and eliminates the need to estimate multicomponent properties solely from binary hydride data. Furthermore, the KKR--CPA framework can be applied to different crystal structures, including BCC, BCT, FCC, and HCP, while retaining its computational advantages. The use of a single global fitting parameter 
yields a particularly good description of the second-plateau (BCT--FCC) widths. For the 22 experimental conditions considered, 95.5\% of the calculated widths fall within a factor of 1.5 of the experimental values, with a mean absolute relative error (MARE) of 26.1\%. These results highlight the ability of the present model to combine ab initio-based thermodynamic description, computational efficiency, and predictive accuracy, making it a promising approach for the systematic screening of multicomponent hydrogen-storage alloys.

\section{Conclusions}

In this work, a thermodynamic model was developed to describe the PCT behavior of multicomponent Ti--V--Nb-based hydrogen-storage alloys. The model combines DFT-derived formation enthalpies obtained within the CPA-KKR framework, a finite-temperature molecular hydrogen reference based on NIST thermodynamic data, and the configurational entropy associated with hydrogen occupation of interstitial sites. Despite its deliberately simple formulation, the model generates the characteristic BCC--BCT--FCC hydrogenation sequence and the corresponding PCT response through Gibbs free-energy minimization and common-tangent constructions.

A central result of this work is that the PCT diagrams emerge from the underlying thermodynamic description rather than being constructed through independent empirical adjustments of the plateau pressures or hydrogen-composition intervals. For each alloy composition and temperature, the equilibrium pressures and coexistence compositions are determined from the calculated free-energy curves. The BCC--BCT and BCT--FCC transitions therefore arise naturally from the relative stability of the corresponding hydrogenated configurations. In particular, the second-plateau pressure and width are simultaneous consequences of the BCT--FCC common-tangent construction.

The entire thermodynamic description requires only one adjustable parameter, $A_{\mathrm{298K}}$, which provides a global energetic alignment of the molecular hydrogen reference. This parameter was optimized simultaneously against the experimental equilibrium pressures and plateau widths of the second plateau over 22 experimental conditions spanning seven alloy compositions. Thus, the model provides a unified description of the PCT behavior across different chemical compositions and temperatures using a single global adjustment.

The global assessment revealed a particularly good reproduction of the second-plateau widths. All 22 calculated widths fall within a factor of two of the experimental estimates, while 95.5\% are within a factor of 1.5. The mean absolute relative error is 26.1\%, with $\mathrm{RMSE}=0.215$ and $\mathrm{MAE}=0.200$. This agreement is noteworthy because the calculated widths are not independently fitted for each experimental condition. Instead, they emerge from the two hydrogen concentrations defining the BCT--FCC common-tangent contacts. The results therefore indicate that the DFT-derived energetic landscape, combined with the configurational entropy, captures an important part of the concentration dependence governing the second two-phase equilibrium.

The agreement in equilibrium pressure is more moderate. The logarithmic pressure errors are $\mathrm{RMSE}_{\log_{10}}=0.759$ and $\mathrm{MAE}_{\log_{10}}=0.686$, with a median multiplicative deviation of 5.36. Although 86.4\% of the calculated pressures fall within one order of magnitude of the experimental values, 22.7\% are within a factor of two and 9.1\% are within a factor of 1.5. These deviations should be interpreted considering the uncertainties associated with both experimental measurements and theoretical approximations. Experimentally, differences between nominal and actual alloy compositions may arise from segregation, incomplete homogenization, synthesis kinetics, diffusion, and characterization uncertainties. Theoretically, the CPA-KKR calculations do not explicitly account for atomic relaxation or hydrogen-induced lattice distortions, while vibrational and anharmonic contributions are also neglected. Furthermore, the exponential dependence of equilibrium pressure on the hydrogen chemical potential means that relatively small differences in this quantity can lead to substantial pressure deviations. Therefore, the observed discrepancies may reflect the combined influence of several sources of uncertainty rather than a single limitation of the model.

Finally, the use of the KKR-CPA framework provides an important computational advantage for investigating chemically disordered multicomponent alloys and their hydrides. By treating substitutional disorder through an effective medium, the method avoids the need for large supercells and explicit atomic configurations, substantially reducing the computational time of evaluating different compositions and hydrogen concentrations. This efficiency makes the present approach particularly promising for high-throughput thermodynamic screening and for generating datasets to support future machine-learning approaches to the discovery and optimization of hydrogen-storage alloys. Although explicit supercell calculations remain valuable for investigating local atomic relaxations and short-range chemical correlations, the combination of CPA-KKR-derived energetics and thermodynamic modeling offers a practical route toward exploring a broader compositional space. Overall, the results demonstrate the potential of this approach to capture important features of the PCT behavior of multicomponent alloys while maintaining a relatively low computational cost.

\section*{Data and Code Availability}



\section*{Acknowledgments}

This work was financed in part by the Conselho Nacional de Desenvolvimento Científico e Tecnológico - CNPq (Grant numbers: 407906/2022-3, 305715/2024-0 and 402122/2025-9), and in part by the São Paulo Research Foundation (FAPESP), Brasil (Process Number 2025/06227-4).



\section*{Funding}



\bibliographystyle{unsrt}
\bibliography{references}

\end{document}